\documentstyle[epsfig,sprocl]{article}

\def\Journal#1#2#3#4{{#1} {\bf #2}, #3 (#4)}

\def\NPB{{\em Nucl. Phys.} B}
\def\PLB{{\em Phys. Lett.}  B}
\def\PRL{\em Phys. Rev. Lett.}
\def\PRD{{\em Phys. Rev.} D}

\def\be{\begin{equation}}
\def\ee{\end{equation}}
\def\bea{\begin{eqnarray}}
\def\eea{\end{eqnarray}}

\begin{document}

\title{ STOCHASTIC DISORIENTED CHIRAL CONDENSATES }

\author{C. GREINER \footnote{Talk presented at the
Institute of Nuclear Theory Program `Probes of Dense Matter in
Ultrarelativistic Heavy Ion Collisions - part: Chiral Dynamics',
Seattle (USA), April 13 - 24, 1998}, Z. XU}

\address{Institut f\"ur Theoretische Physik, Universit\"at Giessen,\\
D-35392 Giessen, Germany \\
  E-mail: carsten.greiner@theo.physik.uni-giessen.de }

\author{T.S. BIRO}

\address{Research Institute for Particle and Nuclear Physics,\\
  H-1525 Budapest, P.O.Box 49, Hungary \\
  E-mail: tsbiro@sunserv.kfki.hu }

\maketitle\abstracts{
Applying a Langevin description of the linear sigma model we investigate
four different scenarios for the evolution of a disoriented chiral
condensate: annealing or quench with
initial conditions governed by effective `light' or physical mass pions.
We present pion number distributions estimated from the zero mode (i.e.
$k=0$-field) component. The best DCC signal is expected for the quench scenario
with initial conditions centered around
$\langle \sigma \rangle \approx 0 $ as would be the case of effective
light `pions' close to the phase transition.
Our investigations support the idea of looking for
DCC formation in individual events.
}

\section{Disoriented Chiral Condensate}

The idea of so called disoriented chiral condensate (DCC)
first appeared in a work of Anselm \cite{1}
but it was made widely known due to Bjorken
\cite{2} and Krzywicki \cite{3}.
This idea has been carried over to the field of
high energy heavy ion collisions as one of the
most interesting suggestions of exotic phenomena \cite{4} as it might
give direct evidence for the chiral phase transition expected to occur
at high energy densities.
Since then many works appeared on various aspects of DCC
formation in heavy-ion collisions
(for a review see \cite{6}
and references listed).
Usually these considerations assume an initial state
at high temperature in which the chiral symmetry is
restored by vanishing collective fields (or equivalently
by vanishing quark condensates).
Independent thermal fluctuations in each isospin direction
of the $O(4)$ sigma-model are present.
This configuration sits on the top of the barrier of
the potential energy at zero temperature, so a sudden
cooling of the system supposedly brings it into an
unstable state. This picture is referred to as
the quenched approximation \cite{4}.
The spontaneous growth and subsequent
decay of these configurations would give rise to large collective fluctuations
in the number of produced neutral pions compared to charged pions,
and thus could provide a mechanism explaining a family of peculiar
cosmic ray events, the Centauros \cite{La80}. A deeper reason for these
strong fluctuations lies in the fact that all pions are assumed to sit in the
same momentum state and the overall wavefunction can carry no isospin
\cite{Gr93}. 

The proposed quench scenario, however, assumes 
that the effective potential governing the evolution
of the long wavelength modes immediately turns
to the classical one at zero temperature.
This is a very drastic assumption as the soft classical
modes completely decouple from the residual thermal
fluctuations at the chiral phase transition temperature
in an ad hoc manner.  

An alternative, the annealing scenario \cite{L8} was suggested by
Gavin and M\"uller. They used the one-loop
effective potential instead of the classical one including thermal
fluctuations. For only moderate expansion in one or more
dimension this view did not lead to a prediction
of a huge DCC signal. 
The smallness of the order parameter at the beginning
of the DCC formation has also been criticized:
if the soft field remains in thermal contact with
the fluctuations giving rise to the one-loop potential,
then one also has to allow for thermal fluctuations
in the initial conditions \cite{L9,L10}.
Quenched initial conditions within the linear sigma model
are statistically unlikely.

The likeliness of an instability leading potentially to a DCC
event during the evolution with a continuous contact 
with the heat-bath of thermal pions was investigated
by us in \cite{PRL}. In the present contribution
we will first review the theoretical ideas behind a classical Langevin
description of the soft fields employed in \cite{PRL} and
then shortly discuss our model for simulating
the evolution of the order parameter and the soft pion fields.
We then state some new numerical results of the simulation
on the evolution of a coherent pionic field considered
by different scenarios.

Our picture of a possible DCC formation in high energy
heavy-ion collisions is as follows:
\\[2mm]
{\large $\bullet $}
in the first stage of the collision ($\tau = 0 \ldots 0.5$ fm/c)
a parton gas is formed with a temperature well over the chiral restoration
point ($T \gg T_c$),
\\[2mm]
{\large $\bullet $}
in the following ($\tau = 1 \ldots 2$ fm/c)
the temperature is around critical ($ T \approx T_c$) and
small chirally disoriented domains (bubbles) of collective
pionic fields start to form together with a thermalized
background of (quasi-)pions,
\\[2mm]
{\large $\bullet $}
at the time ($\tau = 2 \ldots 10$ fm/c) the temperature
drops and the fireball expands rapidly, at most one bigger
DCC domain survives this phase in the heat bath of thermal pions,
\\[2mm]
{\large $\bullet $}
finally ($\tau \ge 10$ fm/c) the fireball breaks off and
 --- in some events --- many soft pions are emitted with
isospin distribution characteristic to DCC.

\section{Description of soft modes within a thermal environment }

One of the recent topics in quantum field theory at finite temperature
or near thermal equilibrium concerns the evolution and behavior of the long
wavelength modes. These modes often lie entirely in the non-perturbative regime.
Therefore solutions of the classical field equations in Minkowski space have
been widely used in recent years to describe long-distance properties
of quantum fields that require a non-perturbative analysis.
The distinction between soft and hard modes separates the energy and
momentum scales as
$gT \ll k_c \ll T$,
allowing for a perturbative resummation of certain
interactions. Soft modes ($k < k_c$) are non-perturbative,
but - this is the hope - can be treated (semi)classically.
Hard modes ($k > k_c$) are treated perturbatively.
A justification of the classical treatment of the long-distance dynamics
of weakly coupled bosonic quantum
fields at high temperature is based on the observation that the average 
thermal amplitude of low-momentum modes is large and
approaches the classical equipartition limit.
The classical field equations
should therefore provide a good approximation for
the dynamics of such highly occupied modes.
However,
the thermodynamics of a classical
field is only defined if
an ultraviolet cut-off $k_c$ is imposed on the momentum {\bf p} such as a
finite lattice spacing $a$.
Many thermodynamical
properties of the classical field depend strongly on the value of the 
cut-off parameter $k_c$ and diverge in the continuum limit $(k_c \to\infty)$.
In a correct semi-classical treatment of the soft modes the hard modes
thus cannot be neglected, but it
should incorporate their influence in a consistent way.

In a recent paper \cite{GMu97} it was shown how to construct
an effective semi-classical action for describing not only the classical
behavior of the long wavelength modes below some given
cutoff $k_c$, but taking into account also perturbatively the interaction among the soft
and hard modes.
Also for studying non-equilibrium dynamics the
separation of soft and hard contributions is of
importance. The resulting effective action \cite{GMu97,GL98a},
$S_{{\rm eff}}[{\rm soft}]$ turns out to be complex,
leading to a stochastic equation of motion for the soft modes.
If the hard modes are already in thermal equilibrium then
the evolution of the soft modes is described by a set
of generalized Langevin equation - the equation of motion
corresponding to the above complex effective action.

We briefly sketch the above following
\cite{GMu97} by considering a scalar field with interaction
$ {\cal L}_{{\rm int}} = \frac{g^2}{4!} \, \tilde{\phi }^4$.
The splitting of the Fourier-components,
$ \tilde{\phi }(p,t) = \phi(p<k_c,t) \, + \, \varphi(p>k_c,t)$,
leads to the following interaction part in the action
\be
 S_{{\rm int}}[\phi,\varphi] = - \int_{t_0}^t \! d^4x \,\,
     \left( \frac{g^2}{4!}\varphi^4 + \frac{g^2}{3!} \left(
     \phi^3\varphi + \frac{3}{2} \phi^2 \varphi^2 
     +\phi\varphi^3 \right) \right).
\ee
By integrating out the `influence' of the hard modes
on the two-loop level, one notices that
Feynman graphs contributing at order ${\cal O}(g^4)$
to the effective action contain
imaginary contributions.
The stochastic equation of motion has the general shape
\be
\Box \phi + \bar{m}^2 \phi + \frac{g^2}{3!} \phi^3
+ \sum_{N=1}^{3} \frac{1}{(2N-1)!} \phi^{N-1}
({\rm Re} \Gamma_{2N} ) \phi^N =
\sum_{N=1}^3 \phi^{N-1}  \xi_N \, .
\ee
Here $\Gamma_{2N}$ denotes the effective contribution with
$2N$ soft legs, $\bar{m}^2$ the resummed Hartree-Fock
self energy (cactus graphs) and $\xi_N$ are associated
noise-variables with a correlation
$ \langle \xi_N \, \xi_N' \rangle = {\rm Im}\Gamma_{2N} $.
When the characteristic time scale in the heat-bath of
hard modes is short compared to the evolution of the
soft order parameter fields, then the Markovian limit has a form
\begin{eqnarray}
{\partial^2\phi\over\partial t^2} + && \left ( {\bf k}^2+ \tilde m_{eff}^2
\right)\phi + \left(
{\tilde g_{eff}^2\over 6} \right) \otimes \phi^3 +
\nonumber \\
&&\mu_{eff} \otimes \phi^5 + \sum_{N=1,2,3} \eta^{(N)}
\dot{\phi } \;
\approx \; \sum_{n=1}^3 \xi_N \otimes \phi^{N-1}\;. \label{Lang3}
\end{eqnarray}
This leads to the familiar form of
plasmon damping ($`\eta `$) and to momentum dependent mass corrections.
The hard modes act as an
environmental heat bath. They also guarantee that the soft modes become, on average,
thermally populated with the same temperature as the heat bath.
The noise terms are related to the plasmon damping in the Markovian approximation
via the simple fluctuation dissipation relation
\be
\langle \xi(t) \xi(t') \rangle = 2T V \eta  \delta(t-t')
\label{FDR}
\ee
in the semiclassical (high temperature) limit.

\section{Langevin description of linear sigma model}

For the description of the evolution of collective pion
and sigma fields in the $O(4)$-model we utilize the
following simplified equations of motion
(taking $k_c \rightarrow 0$)
for the order parameters $\Phi _a =\frac{1}{V} \int
d^3x \,
\phi_a ({\bf x},t)$ in a volume $V$ \cite{PRL}:
\bea
\ddot{\Phi_0} + \left( \frac{D}{\tau} + \eta \right) \dot{\Phi_0}
+ m_T^2\Phi_0 & = & f_{\pi}m_{\pi}^2 + \xi_0, \nonumber \\
\ddot{\Phi_i} + \left( \frac{D}{\tau} + \eta \right) \dot{\Phi_i}
+ m_T^2\Phi_i & = & \xi_i,
\label{eq1}
\eea
with $\Phi_0=\sigma$ and $\Phi_i=(\pi_1,\pi_2,\pi_3)$
being the chiral meson fields and
\bea
\label{mpi}
 m_T^2 & = & \lambda \left( \Phi_0^2 + \sum_i \Phi_i^2 +
\frac{1}{2} T^2 - f_{\pi}^2 \right) + m_{\pi}^2, \\
\label{msig}
 m_L^2 & = & m_T^2 + 2\lambda \left( \Phi_0^2 + \sum_i \Phi_i^2  \right)
\eea
the effective masses. These coupled Langevin equations resemble
in its structure a Ginzburg-Landau description of phase transition.

Here "dots" denote the time derivative with respect to proper time $\tau$,
$\eta$ the effective damping coefficient and $D/\tau$ is
the Raleigh-damping corresponding to a $D$-dimensional scaling
flow of the fireball.
The later assumed (rapid) expansion and cooling results in time dependent temperature
and volume according to
\be
\frac{\dot{T}}{T} + \frac{D}{3\tau} = 0 \, , \qquad \qquad
\frac{\dot{V}}{V} - \frac{D}{\tau}  = 0 \, .
\ee

For the friction coefficient $\eta $
of the $\sigma $ and pion field we take the on-shell
plasmon damping rate for standard $\phi ^4$-theory
generalized to the O(4)-model:
\be
\label{fric1}
\eta \, = \, 2 \gamma _{pl} \, = \,
\frac{9}{16 \pi ^3} \lambda ^2  \frac{T^2}{m} \, f_{Sp} (1-e^{-\frac{m}{T}})
\, \, \, ,
\ee
where $f_{Sp}(x) = -\int_{1}^{x} dt \frac{\ln t}{t-1} $ defines the
Spence function.
We treat the dissipation term in a first step as Markovian
(mainly because of numerical implementation).
Furthermore we take $m=m_{\pi }=140$ MeV for specifying the
damping cofficient $\eta $ in the following.
(We remark that both rather crude approximations can be relaxed \cite{Xu98}.)
In the semiclassical Markovian approximation the noise is effectively 
white and at two-loop level it is Gaussian \cite{GMu97,GL98a} and thus follows
a similar relation like (\ref{FDR}).

Of course, semi-classical Langevin equations may not hold for a strongly
interacting theory as for highly non trivial dispersion relations 
the frequencies of the
long wavelength modes are not necessarily much smaller than the temperature.
Still, when the soft modes become tremendously populated
one can argue that the long wavelength modes
being coherently amplified behave classically \cite{4}.
Aside from a theoretical justification one can  regard the Langevin
equation as a practical tool to study the effect of thermalization
on a subsystem, to sample a large set of possible trajectories
in the evolution, and to address also the question of all thermodynamically
possible initial configurations in a systematic manner.

\begin{figure}[ht]
\centerline{\epsfxsize=12cm \epsfbox{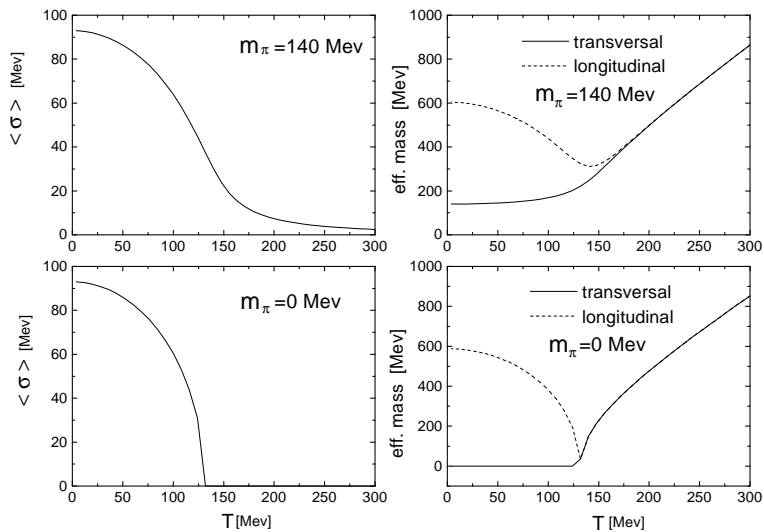}}
\caption{The order parameter $\sigma $, the transverse (eff. 'pion') mass $m_T$
and the longitudinal (eff. 'sigma') mass as a function of temperature $T$
for an equilibrium situation for physical vacuum pion mass and for
vanishing pion mass.
\label{MASS}}
\end{figure}

The `Brownian' motion of the soft field configuration
leads to equipartition of the energy at constant temperature.
In fig. \ref{MASS} we show the effective transversal masses $m_{T }$ (\ref{mpi})
of the pion modes and $m_{L }$ (\ref{msig})
of the $\sigma $ mode as a function of the temperature obtained with
solving eqs. (\ref{eq1}) for a static equilibrium system (i.e. $D=0$)
at fixed
temperature $T$ and sufficiently large volume $V$. The masses shown
are thus taken as an ensemble average of the different realizations
of the Langevin scheme. For large volumes the fluctuations
in the obtained masses are of the order $1/V$ and thus small.
For the situation that the vacuum pion
mass is assumed to be zero (no explicite symmetry breaking)
one can realize from fig. \ref{MASS} the situation for a true second order
phase transition occuring at the transition temperature
$T=T_c\equiv \sqrt{2 f_{\pi}^2 - 2m_{\pi }^2/\lambda } \approx  125$ MeV.
On the other hand for the physical situation of a nonvanishing pion mass
of $m_{\pi}=140 $ MeV the `phase transition' resembles more the form
of a smooth crossover. In this case, at $T \approx T_c$, the
$\sigma $-field still posseses a nonvanishing value of
$\langle \sigma (T\approx T_c)\rangle \approx f_{\pi }/2 \approx \sigma _{vac}/2$.

\section{Stochastic DCC}

In ref. \cite{PRL} we had studied
the average and statistical properties of individual solutions of the
above Langevin equations with the emphasis on such periods of
the time evolution when the transverse mass $m_T$ becomes imaginary
and therefore an exponential growth of unstable fluctations in the
collective fields might be expected.
We had found that in different realistic initial volumes
individual events lead to sometimes significant growth
of fluctuations.

\begin{figure}[ht]
\centerline{\epsfxsize=12cm \epsfbox{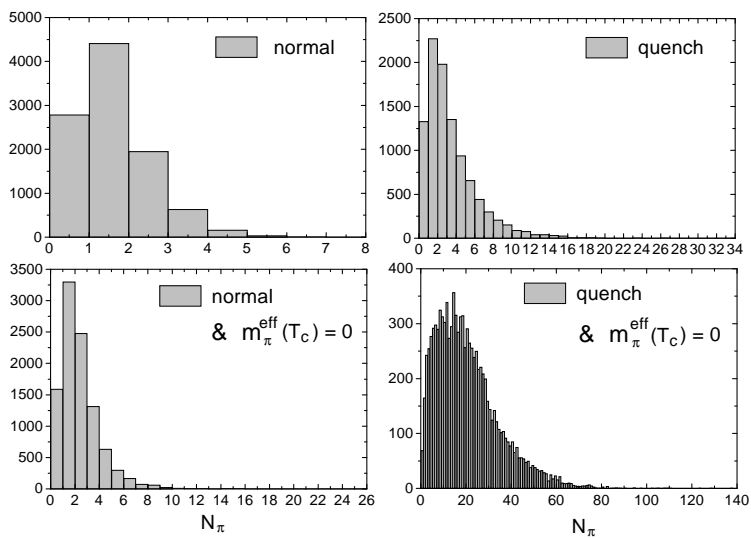}}
\caption{ The pion number distribution
in four different DCC scenarios.
\label{PION}}
\end{figure}

Here we want to show now
the distribution in the number of produced
long wavelength pions $N_\pi $ out of the evolving
chiral pion order fields within the DCC domain $V(\tau )$ \cite{Xu98}.
Within the semiclassical interpretation these can approximately be written as
\be
N_{\pi}^{{\rm zero mode}} \approx \frac{1}{2} m_{\pi}
\left( \vec{\pi }^2 + \frac{1}{m_{\pi}^2} \dot{\vec{\pi }}^2 \right) \, V(\tau ) \, .
\ee
As a rough estimate one can think of
the quench energy density $\frac{\lambda }{4} f_\pi ^4 $ being transformed into
low momentum pions yielding
$N_{\pi}/V \approx \frac{\lambda}{4}\frac{f_{\pi}^4}{m_{\pi}}
\approx 0.3 \, fm^{-3} $. On the other hand one might also estimate that
the chiral order field `circles' around with
$ \langle \vec{\pi }^2 + \frac{1}{m_{\pi}^2} \dot{\vec{\pi }}^2 \rangle \,
\leq f_\pi ^2 $ resulting in  $N_{\pi }/V \approx f_\pi^2/(2m_\pi )
\approx  0.08 \, fm^{-3}$. For the total pion number
the crucial question is then how large has the evolving
volume $V(\tau )$ of the DCC domain increased when the pion oscillations
have emerged after the roll down period.

In the following we present results for four different DCC scenarios:
\\[2mm]
{\large $\bullet $}
{\it Normal} or annealing scenario \cite{PRL} :
The system is preheated at $T=T_c$ for sufficiently long time
in order to cover a complete set of possible initial thermal conditions
using its own dynamics (\ref{eq1}) and then switch over to
a D-dimensional scaling expansion.
\\[2mm]
{\large $\bullet $}
{\it Quench} scenario: As for {\it Normal}, however, after preheating
and thus during the expansion the term  $\lambda /2 T^2$ of the potential in
the equations of motion (\ref{eq1},\ref{mpi}) is being omitted in order to mimique
an abrupt occurence of the zero temperature vacuum potential.
\\[2mm]
{\large $\bullet $}
$m_{\pi }^{eff} (T=T_c) \approx 0$ scenario:
As for {\it Normal}, however, the preheating at $T=T_c$ is carried out with
$m_\pi = 0$ (comp. fig. 1) in order to generate initial conditions
which are centered around the origin $\sigma = \vec{\pi } =0$.
\\[2mm]
{\large $\bullet $}
{\it Quench} and $m_{\pi }^{eff} (T=T_c) \approx 0$ scenario:
Like the previous, however, the
$\lambda /2 T^2$ term of the potential in
the equations of motion (\ref{eq1}) is being omitted.

One might argue that the use of the initial conditions prepared within the
$m_{\pi }^{eff} (T=T_c) \approx 0$ scenario is inconsistent within the
linear sigma model with a physical pion mass. From fig. 1
one notices that then the phase transition resembles a smooth crossover.
From lattice calculations, however, one knows that the chiral transitions
happens much sharper within a very narrow window close at $T=T_c$.
This means that it might very well be that the order parameter
will fluctuate around zero slightly above the critical temperature.

In fig. 2 we depict the distribution of produced pions within
$10^4$ events within the four different DCC scenarios.
As parameters we had taken\cite{L8} $D=3$, $\tau_0 =7\, fm/c$ and
$V(\tau_0 )=100 \, fm^3$.
A comparison of annealing and quench scenarios, both
with finite and vanishing pion mass (for generating the {\em initial} conditions)
reveals that
the most productive DCC events would lead to a few (6-8 in
annealing scenario with finite pion mass), to a moderate number
(20 - 40 in annealing scenario with massless pions and quench with
massive pions) or to about 120 pions (in quench scenario
with initial conditions generated by massless pions), respectively.
The final distribution does not follow a usual Poissonian distribution.
Instead, fluctuations with a large number of produced pions are still likely
with some finite probability! In principle, an ensemble averaged
description of potential DCC formation carried out
within the mean field approximation,
as presented in the various literature, can not account
for such fluctuations and thus has to fail at some point.
Also the isospin ratio signal
in the latter three scenarios is close to that of expected
for a DCC event.
These findings support the idea of looking for
DCC formation in individual events.

\begin{figure}[ht]
\centerline{\epsfxsize=12cm \epsfbox{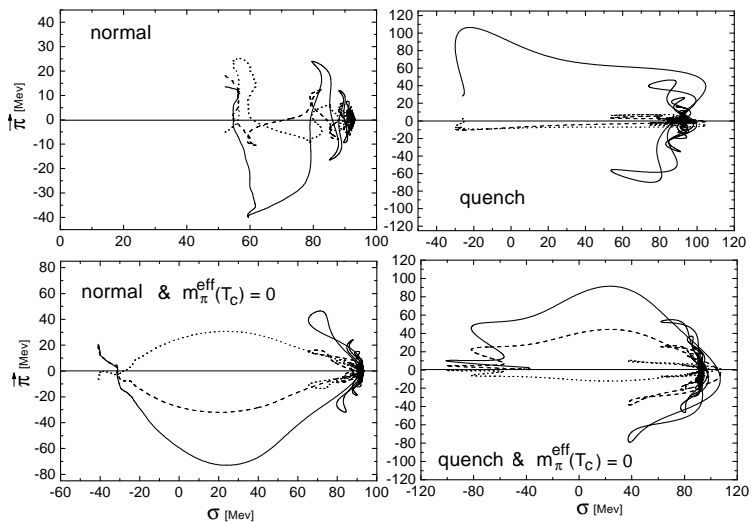}}
\caption{ The trajectories of the three soft pion modes $\pi ^i$
in the $\pi \, - \, \sigma $-plane for each of the particular events
with the largest pion number out of fig. 2
in the four different DCC scenarios.
\label{Traject}}
\end{figure}

In fig. 3 we depict
the individual trajectories of the best candidates.
Especially for the first scenario one recognizes the fact that the initial
conditions for the order parameters are centered around $\sigma \approx f_\pi /2$
(comp. with fig. 1) and do not allow for initial fluctuations in the backward
hemisphere with $\sigma <0$.
In any case, all these trajectories clearly illustrate that
the collective pion field in the late stage of the evolution within the
scaling expansion oscillates around the `chiral' circle
$\sigma ^2 + \vec{\pi }^2 \approx f_\pi^2 $ ($f_\pi = 93 $ MeV) as expected
for DCC formation.

\section{Summary and Conclusion}

In summary we want to remark that within
the presented Langevin description
one can simulate,
on an event by event analysis,
the possible evolution of various DCC scenarios in a rather transparent form.
It has been demonstrated that although the average evolution of
the order parameter field is damped, there are individual events
which show pionic instabilities and
appreciable pion yields.
One can repeat the presented calculations for different sets of parameters
(the expansion dimension $D$, the scaling parameter of inital time
$\tau_0$ \cite{Ra96a} and the initial volume $V(\tau _0$). In addition, one can also start to describe the evolution
with initial conditions generated at much higer initial temperatures
$T \gg T_c$ as carried out by Randrup \cite{Ra96a}. We will leave this for
future work \cite{Xu98}.

\section*{Acknowledgments}
This work was supported by the Deutsche Forschungsgemeinschaft
(DFG) and the Hungarian Academy of Sciences (MTA) (Project. No 101)
and by the Hungarian National Research Fund (OTKA)
(Project No. T019700, T024094).

\end{document}